\documentclass[fleqn,usenatbib]{mnras}
\usepackage{newtxtext,newtxmath}
\usepackage[T1]{fontenc}

\DeclareRobustCommand{\VAN}[3]{#2}
\let\VANthebibliography\thebibliography
\def\thebibliography{\DeclareRobustCommand{\VAN}[3]{##3}\VANthebibliography}

\usepackage{mhchem}
\usepackage{leftindex}
\usepackage[colorinlistoftodos]{todonotes}

\usepackage{amssymb}
\usepackage{comment}
\usepackage{multirow}
\usepackage{listings}
\usepackage{makecell}
\usepackage{booktabs}
\usepackage{amsmath}
\usepackage{subfigure}
\usepackage{xcolor}
\usepackage{hyperref}
\usepackage{graphicx}
\usepackage{bm}
\usepackage{float}
\usepackage{verbatim}
\usepackage{enumitem}
\usepackage[linesnumbered,ruled]{algorithm2e}

\hypersetup{colorlinks=true, citecolor=blue,
  linkcolor=cyan, urlcolor=magenta}

\usepackage{color}
\newcommand{\proptosim}{\mathrel{\vcenter{
 \offinterlineskip\halign{\hfil$##$\cr
 \propto\cr\noalign{\kern2pt}\sim\cr\noalign{\kern-2pt}}}}}

\renewcommand{\d}{\mathrm{d}}

\newcommand{\e}{\mathrm{e}}

\renewcommand{\P}{\mathrm{P}}     

\newcommand{\kratos}{Kratos-polrad}
\renewcommand{\Im}[1]{\ensuremath{{\rm Im}[#1]}}
\renewcommand{\Re}[1]{\ensuremath{{\rm Re}[#1]}}

\newcommand{\figdir}{.}

\title[Kratos-polrad: GPU Monte-Carlo polarized radiative transfer]
{Kratos-polrad: Novel GPU system for Monte-Carlo simulations
  with consistent polarization calculations}

\author[Yang and Wang]{
Haifeng Yang,$^{1,2}$\thanks{E-mail: hfyang@zju.edu.cn}\thanks{ZJU Tang Scholar}
Lile Wang,$^{3,4}$\thanks{E-mail: lilew@pku.edu.cn}
\\
$^{1}$Institute for Astronomy, School of Physics,
  Zhejiang University, 886 Yuhangtang Road, Hangzhou 310027, China\\
$^{2}$Center for Cosmology and Computational
  Astrophysics, Institute for Advanced Study in Physics,
  Zhejiang University, Hangzhou 310027, China\\
$^{3}$The Kavli Institute for Astronomy and
  Astrophysics, Peking University, Beijing 100871, China\\
$^{4}$Department of Astronomy, School of Physics,
  Peking University, Beijing 100871, China
}

\date{Accepted XXX. Received YYY; in original form ZZZ}

\pubyear{\the\year}

\begin{document}
\label{firstpage}
\pagerange{\pageref{firstpage}--\pageref{lastpage}}
\maketitle

\begin{abstract}
  Polarized radiation serves as a vital diagnostic tool in
  astrophysics, providing unique insights into magnetic
  field geometries, scattering processes, and
  three-dimensional structures in diverse astrophysical
  scenarios. To address these applications, we present
  \kratos{}, a novel GPU-accelerated Monte Carlo Radiative
  Transfer code built upon the heterogeneous computing
  framework of Kratos, designed for self-consistent and
  efficient polarization calculations.  It utilizes a
  comprehensive treatment of the Stokes parameters throughout
  photon propagation, featuring grain--lab frame
  transformations using quaternion algebra and a consistent
  analytical integration of polarized extinction and emission
  within each cell, which are useful in modeling radiative
  transfer processes with scatterings by aligned dust grains. The code implements
  two-step polarimetry imaging that decouples Monte Carlo
  sampling of scattering physics from imaging geometry,
  enabling efficient synthesis maximizing the utilization of
  photon packets.  Extensive validation against analytical
  solutions and established codes demonstrates accurate
  treatment of diverse polarization phenomena, including
  self-scattering polarization, dichroic extinction in
  aligned dust grains, and complex polarization patterns in
  twisted magnetic field configurations. By leveraging
  massive GPU parallelism, optimized memory access patterns,
  and analytical approaches for optically thick cells,
  \kratos{} achieves performance improvements of
  $\sim 10^{2}$ times compared to CPU-based methods,
  enabling previously prohibitive studies in polarimetric
  astrophysics.
\end{abstract}


\section{Introduction}
\label{sec:intro}

Polarimetric observations at millimeter and near-infrared
wavelengths provide unique diagnostics of dust grain
properties, magnetic field geometries, and alignment
mechanisms in protoplanetary disks and star-forming
regions. At millimeter wavelengths accessible to ALMA,
polarization in many disks shows a characteristic
wavelength-dependent transition: self-scattering dominates
at shorter wavelengths (e.g., Band~7;
\citealt{2015ApJ...809...78K,Yang2016a,2017ApJ...851...55S,2018ApJ...860...82H,2018ApJ...865L..12B,2019MNRAS.482L..29D,2024ApJ...963..134Y,2026A&A...708A..95L}),
while dichroic emission from aligned grains takes over at
longer wavelengths (e.g., Band~3;
\citealt{2017ApJ...851...55S,2017ApJ...844L...5K,2024ApJ...967...40H}),
with recent panchromatic observations tracing
this transition down to centimetre wavelengths
\citep{Lin2024a}.  In the near-infrared,
high-contrast polarimetry with VLT/SPHERE and Gemini/GPI
\citep{2018A&A...619A...9S,2018ApJ...863...44A,2022AJ....164..109R,2023A&A...680A.114R,2024A&A...681A..19C} probes scattered starlight from
disk surfaces; when grains are aligned, the observed
polarization deviates systematically from the azimuthal
pattern expected for spherical scatterers, with
position-angle offsets reaching $\sim 10^\circ$ or more
\citep{Yang2022}.  Both regimes converge on a single
unresolved problem: the self-consistent treatment of
scattering by aligned, non-spherical dust grains in Monte
Carlo radiative transfer.

Modeling polarized radiation in these environments requires
solving the vector radiative transfer equation for the full
Stokes parameters $(I, Q, U, V)$
\citep{1960ratr.book.....C}.  Monte Carlo radiative
transfer (MCRT) \citep{2013ApJS..207...30W} has become the
standard approach for three-dimensional geometries, and
several well-established codes include polarization
capabilities: RADMC-3D \citep{2012ascl.soft02015D}, MCFOST
\citep{2006A&A...459..797P}, POLARIS
\citep{2016A&A...593A..87R}, SKIRT
\citep{2015A&C.....9...20C, 2020A&C....3100381C}, and
Hyperion \citep{2011A&A...536A..79R}.  Among these, POLARIS
treats aligned-grain dichroic emission and spherical-grain
scattering as separate processes; the remaining codes are
restricted to spherical grains.  Only the HOCHUNK code of
\citet{2002ApJ...574..205W} has self-consistently coupled
scattering with grain alignment, and no modern, publicly
maintained code currently provides this capability.  Yet
exactly this coupling is essential wherever aligned grains
both scatter and emit polarized radiation,precisely the
regime now probed by ALMA and SPHERE observations of
protoplanetary disks.

The physical origin of mm-wavelength polarization in disks
is actively debated.  While interstellar grains align with
magnetic fields through radiative torques
\citep{2007MNRAS.378..910L, 2015ARA&A..53..501A},
\citet{Yang2019} showed that the azimuthal polarization
of HL~Tau
\citep{2017ApJ...844L...5K} is better explained by prolate
grains aerodynamically aligned through gas-dust drift,
rather than the radiative alignment scenario of
\citet{2017ApJ...839...56T}.
\citet{2024ApJ...967...40H} found similar multi-wavelength
transitions in other disks, with multiple polarization
origins often coexisting, while
\citet{2025NatAs...9..526O} identified magnetic grain
alignment in the southern region of HD~142527 alongside
self-scattering in the north, showing that distinct
mechanisms can operate within a single disk.
\citet{Lin2024b} provided a theoretical foundation through
``birdie-like'' aerodynamic grain alignment.  Only
\citet{2024ApJ...970..114T} have argued for magnetic
alignment, adopting ``wrong'' internal grain alignment
(long axes parallel to angular momentum).
These competing mechanisms , radiative, aerodynamic,
and magnetic , directly modulate the scattering Mueller
matrix and dichroic extinction cross-section, yet the
dominant alignment is unknown
{\it a priori} and likely varies with disk conditions.
Disentangling the contributions of aligned-grain scattering
and dichroic emission to the observed polarization is
therefore essential for correctly inferring grain sizes,
compositions, and the underlying alignment physics.

\setcounter{footnote}{0}

To address this gap, we present \kratos{}\footnote{\url{https://github.com/wll745881210/kratos_polrad.git}}, a
GPU-accelerated polarized radiative transfer module built
upon the Kratos heterogeneous-computing framework
\citep{2025ApJS..277...63W}.  Kratos provides a device
abstraction layer for heterogeneous architectures,
mesh-management infrastructure, and mixed-precision
numerical methods on consumer-grade GPUs.  \kratos{} extends
this framework to polarized MCRT with aligned grains,
introducing quaternion algebra , a novel approach in
astronomical MCRT , for numerically robust Stokes-frame
coordinate transformations, and is the first publicly
available code to self-consistently couple aligned-grain
scattering with dichroic emission and extinction.

\section{Polarized Radiative Transfer with Quaternion Coordinates}
\label{sec:methods}

\kratos{} employs a GPU-accelerated Monte Carlo framework
built upon the Kratos simulation system
\citep{2025ApJS..277...63W}, utilizing the grid-based
ray-tracing method described in
\citet{2025arXiv250404941W}. A composite mesh grid,
optionally enhanced with static mesh refinement, stores the
physical state variables. Photon packets propagate through
this grid, undergoing emission, absorption along their path,
and scattering events that redirect their
trajectories. However, polarization-aware calculations
require additional treatment: the Stokes parameters carried
by each photon packet must be transformed via Mueller
matrices at every scattering interaction, and the
accumulated polarization state must be tracked throughout
the photon's lifecycle to enable synthesis of polarized
observables.

For spherical dust grains, the coordinate transformations
are straightforward: the grain optical properties are
azimuthally symmetric around the scattering direction, so
the reference frame is naturally aligned with the photon
propagation geometry.  For aligned, non-spherical grains,
however, all optical properties , the scattering Mueller
matrix and the dichroic extinction cross-section , are
defined in the grain-aligned frame, whose orientation is
set by the local alignment axis and generally does not
coincide with the photon propagation direction.
Transformations between the laboratory frame and the
grain-aligned frame are therefore required at every
scattering and propagation step, and these rotations are
prone to the gimbal lock problem when implemented with
Euler angles.  We transform the photon state into
the grain frame at each cell, rather than absorbing the
basis rotations into an effective Mueller operator in a
fixed global basis, because the extinction matrix and the
scattering Mueller matrix are defined in the grain frame;
realistic optical properties from sophisticated methods,
such as the T-matrix method \citep{1998JQSRT..60..309M} or
the discrete-dipole approximation (DDA)
\citep{1994JOSAA..11.1491D}, are likewise naturally computed
there.  This motivates the
use of quaternion algebra described below.

\subsection{Quaternion-based coordinate transformations for
  polarized radiative transfer}
\label{sec:method-quat}

We employ Hamilton's quaternions, which provide a compact
representation of rotations that enables stable, efficient
composition of successive transformations directly in GPU
registers. A quaternion
$\mathbf{q} = q_0 + q_1\mathbf{i} + q_2\mathbf{j} +
q_3\mathbf{k}$ extends complex numbers into four dimensions,
where the basis elements satisfy the fundamental relations:
\begin{equation}
\mathbf{i}^2 = \mathbf{j}^2 = \mathbf{k}^2 = \mathbf{ijk} = -1.
\end{equation}
This algebraic structure provides a natural representation
for three-dimensional rotations without the singularities
inherent in other parameterizations.

In polarized radiative transfer through aligned dust grains,
the local scattering frame (the ``Grain Frame'') is
determined by the symmetry axis of the dust particles. For
this work, we assume perfect alignment with the local
magnetic field direction, characterized by position angles
$(\theta_B, \phi_B)$ in global spherical coordinates. The
rotation from the global frame to the Grain Frame is
represented by the quaternion:
\begin{equation}
\label{eq:grain-quaternion}
\mathbf{q}_\mathrm{B} =
\begin{cases}
q_{B,0} = \ \ \cos(\phi_B/2) \cos(\theta_B/2)\\
q_{B,1} = -\sin(\phi_B/2) \sin(\theta_B/2)\\
q_{B,2} = \ \ \cos(\phi_B/2) \sin(\theta_B/2)\\
q_{B,3} = \ \ \sin(\phi_B/2) \cos(\theta_B/2)
\end{cases},
\end{equation}
which follows the convention established in the
\texttt{quaternion} Python module \citep{quaternion}. This
specific form ensures proper handling of the double coverage
of the rotation group by quaternions.

For a photon propagating along direction $\hat{n}$ in the
global frame, its representation in the Grain Frame
$\hat{n}_{GF}$ is obtained through the quaternion
conjugation:
\begin{equation}
\label{eq:vector-transform}
(0, \vec{n}_{GF}) = \mathbf{q}_\mathrm{B}^*
\cdot (0, \vec{n}) \cdot \mathbf{q}_\mathrm{B},
\end{equation}
where $(0, \vec{n})$ denotes a pure quaternion (zero scalar
component) constructed from the vector components, and
$\mathbf{q}_\mathrm{B}^*$ represents the quaternion
conjugate. This operation efficiently rotates the
propagation direction into the local grain-aligned frame,
which determines the incident direction for scattering
calculations.

The transformation of polarization states requires careful
treatment of the reference direction conventions. We adopt
a convention where for light
propagating along direction $(\theta, \phi)$, the Stokes
parameters are defined relative to the spherical coordinate
basis vectors:
\begin{align}
\hat{\theta} &= (\cos\theta\cos\phi, \cos\theta\sin\phi,
               -\sin\theta), \\
\hat{\phi} &= (-\sin\phi, \cos\phi, 0).
\end{align}
In this convention, positive Stokes $Q$ corresponds to
linear polarization along the $\hat{\theta}$ direction,
while negative $Q$ indicates polarization along
$\hat{\phi}$. When rotating to the Grain Frame, the
reference direction $\hat{\phi}$ from the global frame
transforms to $\hat{\phi}_\mathrm{GF}$ in the Grain
Frame. However, this differs from the natural $\hat{\phi}'$
direction defined by the rotated propagation vector
$\hat{n}_\mathrm{GF}$. The angle $\eta$ between these two
directions is computed via the cross product:
\begin{equation}
\label{eq:rotation-angle}
\sin\eta = (\hat{\phi}_\mathrm{GF} \times \hat{\phi}')\cdot\hat{n}_\mathrm{GF},
\quad \cos\eta = \hat{\phi}_\mathrm{GF} \cdot \hat{\phi}'.
\end{equation}
This angle $\eta$ then rotates the linear polarization
components according to:
\begin{equation}
\label{eq:stokes-rotation}
\begin{pmatrix}
Q' \\
U'
\end{pmatrix}
=
\begin{pmatrix}
\cos 2\eta & \sin 2\eta \\
-\sin 2\eta & \cos 2\eta
\end{pmatrix}
\begin{pmatrix}
Q \\
U
\end{pmatrix},
\end{equation}
while Stokes $I$ and $V$ remain invariant under this
reference frame rotation.

To validate the coordinate transformation
implementation, we performed a round-trip test: photons with
randomized Stokes parameters and propagation directions were
transformed into the Grain Frame (determined by randomly
oriented magnetic fields) and then transformed back to the
Lab Frame.  Across $10^7$ independent trials, all four
Stokes parameters were recovered to machine precision, 
confirming that the
quaternion-based transformations are unitary and
numerically lossless.  This validation anchors the
credibility of all coordinate-dependent calculations
throughout the remainder of this work.
A complementary test under spatially varying alignment
fields , where light propagates through a medium with a
continuously twisting grain-aligned frame , is presented
in \S~\ref{sec:twisted-bfield-test}.

\subsection{Radiative transfer equations}
\label{sec:method-rt}

The propagation of polarized radiation through astrophysical
media is governed by the vector radiative transfer equation,
which extends the scalar form to track the full Stokes
vector $\mathcal{I} \equiv (I, Q, U, V)^T$. For each frequency
bin (suppressed for notational clarity), the transfer
equation reads:
\begin{equation}
  \label{eq:polarized-rt}
  \dfrac{\d \mathcal{I} (\Omega, x)}{\d \tau} =
  - \alpha \mathcal{I}(\Omega, x )
  + \alpha_{\rm abs} B(x) + S(\Omega, x)\ ,
\end{equation}
where $\Omega$ denotes the photon propagation direction, $x$
the spatial coordinate, $\alpha$ and $\alpha_{\rm abs}$ the
dimensionless extinction and absorption
coefficient matrices respectively, and $B$ the blackbody
emission source term in the Stokes vector form [proportional to
$(1,0,0,0)$ without polarization].
The differential extinction optical depth $\d\tau$
is related to the distance $\d s$ traveled along the
propagation direction by
$\d\tau = \lambda_{\rm ext}^{-1}\d s$, where the reciprocal of
extinction mean free path (MFP) is the sum of scattering and
absorption components
($\lambda^{-1}_{\rm ext} = \lambda^{-1}_{\rm abs} +
\lambda^{-1}_{\rm sca}$), which are related to the total
cross sections via,
\begin{equation}
  \label{eq:mfp-matrices}
  \lambda^{-1}_{\rm abs} = \sum_i n_i \sigma_{{\rm abs},i}\ ,
  \quad
  \lambda^{-1}_{\rm sca} = \sum_i n_i \sigma_{{\rm sca},i}\ ,
\end{equation}
where $n_i$ is the number density of the $i$th component that
participates radiation-matter interactions, and $\sigma_{\rm abs}$
and $\sigma_{\rm sca}$ are the total
cross sections of absorption and scattering, respectively. In
practice, \kratos{} reads in both $\lambda^{-1}_{\rm abs}$ and
$\lambda_{\rm sca}^{-1}$ spatial distribution profiles from input
binary files, and the dimensionless $\alpha$ and $\alpha_{\rm abs}$
matrices are calculated through every step of photon propagation.
The scattering source term $S$ describes how incident
radiation from all directions is redistributed into
direction $\Omega$,
\begin{equation}
  \label{eq:scattering-source}
  S (\Omega) = \oint \d\Omega'\ Z(\Omega; \Omega')
  \mathcal{I}(\Omega')\ ,
\end{equation}
where $Z(\Omega;\Omega')$ is the dimensionless
scattering Mueller matrix.
For anisotropic scattering by aligned dust grains or
electrons,
$Z(\Omega;\Omega')$ depends on both the
scattering angle and the local magnetic field orientation,
requiring careful treatment of coordinate transformations.

Similar to established Monte Carlo radiative transfer codes
(see e.g. \citealt{2013ARA&A..51...63S} and references therein),
\kratos{} numerically samples the polarized radiative
transfer equations (eqs.~\ref{eq:polarized-rt} and
\ref{eq:scattering-source}) by propagating discrete photon
packets through the computational domain, and locating the
scattering events according to a scattering optical depth
obeying exponential distribution generated stochastically at
the moments when photons are generated or deflected by the
previous scattering event. At each scattering event, the
final direction is determined by
sampling the polarization-aware differential scattering
cross section , the first element of
$Z(\Omega; \Omega')\, \mathcal{I}$ in
eq.~\ref{eq:scattering-source} , which encodes the
angular and polarization-dependent scattering properties
of the dust grains at the photon wavelength.
The post-scattering photon properties, especially the Stokes
parameters, are determined with the Mueller matrix in the
coordinate frames of dust grains, where the coordinate transforms
are conducted
using the scheme described in \S\ref{sec:method-quat}. The
differential cross section and Mueller matrix could be defined
by the user prior to the execution of the program, either via
analytical approximations, interpolation tables, or even artificial
neural networks to be included in future works. Between
scattering events, these packets traverse the simulation
mesh from cell interface to cell interface, with each
containing the necessary physical properties to model
radiation-matter interactions
\citep{2025arXiv250404941W}.

The integration of photon packet extinction through each
cell requires care when the extinction optical depth across a
cell, $\delta\tau$, is not small, because a first-order
expansion of the formal solution in $\delta\tau$ then becomes
inaccurate. This is a limitation of the cell-by-cell
integration, not a non-linearity of the transfer equation,
which remains linear in the Stokes vector for fixed
coefficients. \kratos{} therefore integrates the transfer
equation analytically within each
cell (Appendix~\ref{sec:apdx-ana-pol-rt}), retaining
the full exponential solution, valid for arbitrarily large
$\delta\tau$. The Monte Carlo sampling of interaction optical
depths already uses the full exponential distribution and
does not assume optically thin cells, so this first-order
approximation would otherwise enter only the deterministic
cell integration.  The analytical
treatment properly accounts for the coupled evolution of all
four Stokes parameters, including the generation of circular
polarization through differential phase delays and the
transformation of polarization states through successive
interactions. This approach ensures accurate evolution of
the photon packets' Stokes parameters through regions of
high optical depth, maintaining numerical stability and
physical consistency even when traversing strongly absorbing
and emitting media.

\subsection{Efficient two-step imaging scheme}
\label{sec:method-imaging}

For direct comparisons with astronomical observations,
synthetic polarimetry images are demanded. A naive imaging
method based on Monte Carlo simulations would simply record
photons that happen to exit the domain within the camera's
solid angle. This approach, however, is proved to be
computationally prohibitive because (1) the camera subtends
an extremely small fraction of $4\pi$ steradians, yielding
minuscule photon counts; (2) multiple scattering events
randomize photon directions, making direct imaging photons
rare even in optically thick environments. Consequently,
generating polarization maps without being limited by shot
noise would require evolving prohibitive number of photon
packets at a given viewing angle.

\kratos{} circumvents this inefficiency issue through a
two-step approach that decouples scattering physics from
imaging geometry. The method exploits the linearity of the
radiative transfer equation and treats scattered photons as
volumetric sources whose contribution to any viewing angle
can be computed {\it a posteriori}. First, \kratos{} evolves
photon packets through the domain using the standard Monte
Carlo procedure, but with a modification: during
propagation, each path integration segment (i.e., not only
at scattering events) contributes to a scattering source
function stored on the grid,
\begin{equation}
  \label{eq:scat-disc-sum}
  S(\Omega_{\rm cam}, x) = \sum_{\rm pp}
  S_{\rm pp}(\Omega_{\rm cam}, x)\ .
\end{equation}
Here $\Omega_{\rm cam}$ is fixed to the camera direction of
interest, the subscript ``pp'' stands for ``photon packet''.
Note that this accumulation occurs not just at
discrete scattering events but continuously along the ray
path, ensuring that single-scattering contributions are also
captured. The summation runs through all photon packets,
each carrying a Stokes vector $\mathcal{I}_{\rm pp}$. In the
discretized Monte Carlo scheme, the integral over solid
angle becomes a summation over all photon packets traversing
a computational cell. The contribution of each photon
packet follows
\begin{equation}
  \label{eq:scat-pp}
  S_{\rm pp}(\Omega_{\rm cam}, x_c) =
  \frac{\delta l_{{\rm pp},c}}{V_c}\,
  Z(\Omega_{\rm cam}; \Omega_{\rm pp})\,
  \mathcal{I}_{\rm pp}\,
  \frac{1-\e^{-\delta\tau_{{\rm pp},c}}}
       {\delta\tau_{{\rm pp},c}}\ ,
\end{equation}
where $\Omega_{\rm pp}$ is the packet propagation direction,
$\delta l_{{\rm pp},c}$ the path length through cell $c$,
$V_c$ the cell volume, and $\delta\tau_{{\rm pp},c}$ the
extinction optical depth along the segment.
Note that this relation already includes the integration of
incident directions of radiation, which is reduced because
the angular distribution of the intensity of each photon
packet is a Dirac delta function along its current
propagation direction.

Photon packets carry photon count per unit time as weights,
namely ``luminosity weights''. For a stellar source, each packet
represents an equal share of the source luminosity; for dust thermal
emission, each cell is assigned a number of packets proportional to
its emitted luminosity, and that luminosity is divided equally among
them, so that the packets originating from a given cell share the same
weight and their sum reproduces the cell emission. A minimum number
of packets is enforced in every cell, so that even faint cells are
not left unsampled; when this floor is active, the luminosity of a
faint cell is spread over more, correspondingly lighter packets,
which preserves the total emission while reducing the shot noise in
faint regions. This floor also determines the smallest photon number
for which the emission from the entire grid is represented, and it is
the reason for the minimum packet number adopted in the
convergence test of \S\ref{sec:convergence}. The scattering matrix $Z$ carries the
scattering normalization, its phase function being normalized
so that $\int\Phi\,\d\Omega=1$, while the single-scattering
albedo $\omega$ enters through the ratio of scattering to
extinction opacity in the second, ray-tracing step. Because
the division by the number of packets is contained in this
weight, no separate $1/N$ factor appears in
Eq.~\eqref{eq:scat-disc-sum}.

We note that the scattering source function is accumulated
for one camera direction in a single Monte Carlo realization; because
the contribution of each photon packet depends on the camera direction
through the scattering matrix, changing the viewing direction requires
repeating the scattering simulation. The method can, however, be
readily extended to accumulate the source function toward several
camera directions within one realization, since the photon propagation
itself is independent of the camera direction.

Second, with $S(\Omega_{\rm cam}, x)$ properly sampled,
\kratos{} perform a deterministic ray-tracing step by
integrating eq.~\eqref{eq:polarized-rt} along lines of sight
toward the camera. When discretized, after traveling
$\delta l$ in each cell, the integration steps are performed
using ray-marching algorithms that runs through the grid
structures, with the same cell-by-cell analytic integration
of the Stokes parameters described in \S\ref{sec:method-rt}
and Appendix~\ref{sec:apdx-ana-pol-rt}.
The second step in total significantly accelerates the
imaging procedure, as the ray tracing is computationally
much cheaper compared to the direct Monte Carlo imaging
simulation whose most outgoing photons are discarded.

On GPUs, kratos-polrad assigns one 
thread per photon packet, and hardware-optimized atomic 
operations are utilized to avoid conflicts in memory 
manipulations in parallel. The data related to geometries, 
extinction intensities, and polarization properties, are 
cached for each photon during the detailed calculations. 
When conducting imaging using
this architecture, \kratos{} could achieve a $10^8$ photon
sampling of the scattering source function within $\sim 7$
seconds on a single RTX 5090 GPU (see also
\S\ref{sec:self-scattering}).  The resulting speedup of
$\sim 10^2$ relative to traditional CPU-based polarized MCRT
codes makes parameter-space exploration and high-resolution
imaging synthesis feasible.

\section{Verifications}
\label{sec:verification}

This section presents a comprehensive suite of tests that
validate the code against analytical solutions and
established benchmarks, while simultaneously demonstrating
the new polarization phenomenology that emerges from
self-consistent treatment of scattering by aligned
grains. Unless specifically noted, we adopt the dipole
approximation for dust grain optical properties and
rejection sampling of scattering differential cross sections
(see also \citealt{2002ApJ...574..205W}) for the simplicity
and clarity of comparisons.  \S\ref{sec:self-scattering}
also presents performance benchmarks quantifying the GPU
acceleration achieved by \kratos{}.

\subsection{Validation of Self-Scattering Polarization}
\label{sec:self-scattering}

\begin{figure}
  \centering
  \includegraphics[width=\linewidth]{\figdir/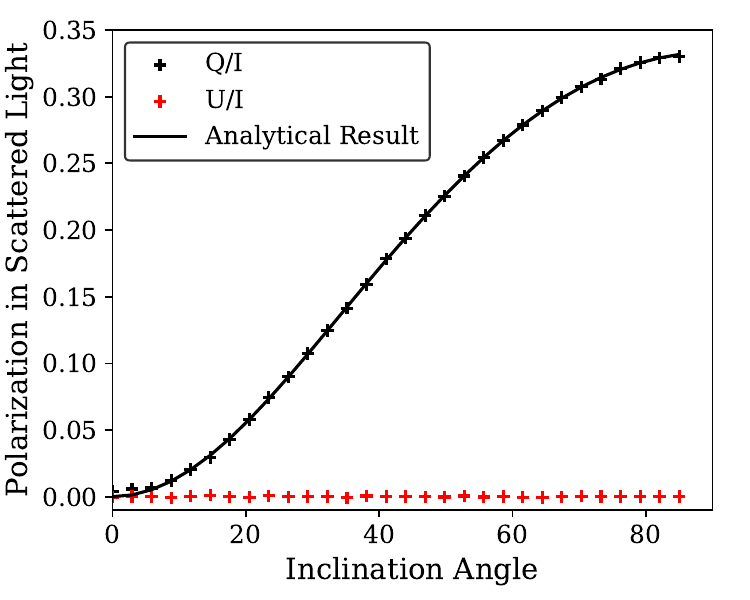}
  \caption{Validation of inclination-induced polarization in
    scattered light. Simulation results for $Q/I$ (black
    markers) and $U/I$ (red markers) are compared against
    the analytical solution (black curve). The agreement
    demonstrates correct implementation of scattering
    polarization dependencies.}
  \label{fig:selfscattering}
\end{figure}

We validate the treatment of self-scattering polarization
through two complementary tests that probe different aspects
of the radiative transfer implementation.  The first test
examines polarization generated purely by scattering in an
inclined disk system. For photons initially emitted within a
thin disk and scattered at oblique angles, the theoretical
polarization fraction follows the analytical
relation\citep{Yang2016a}:
\begin{equation}
    p = \frac{\sin^2 i}{2 + \sin^2 i},
\end{equation}
where $i$ denotes the disk inclination angle. This
expression arises from the differential scattering
cross-sections for polarized light and serves as a
fundamental test of the scattering matrix implementation.
Figure~\ref{fig:selfscattering} compares our simulation
results with this analytical prediction. The normalized
Stokes parameter $Q/I$ closely follows the theoretical curve
across all inclination angles, while $U/I$ remains
consistent with zero (within numerical fluctuations),
confirming the proper orientation of the polarization
vectors relative to the scattering plane.  Note that in
\kratos{}, we always define a local Grain Frame, which has
its symmetric axis along x-direction in this test, even
though the dust grains are in isotropic Rayleigh scattering
regime. The good agreement confirms that \kratos{} can
handle isotropic spherical dust grains very well.

\begin{figure*}
  \centering
  \includegraphics[width=\linewidth]{\figdir/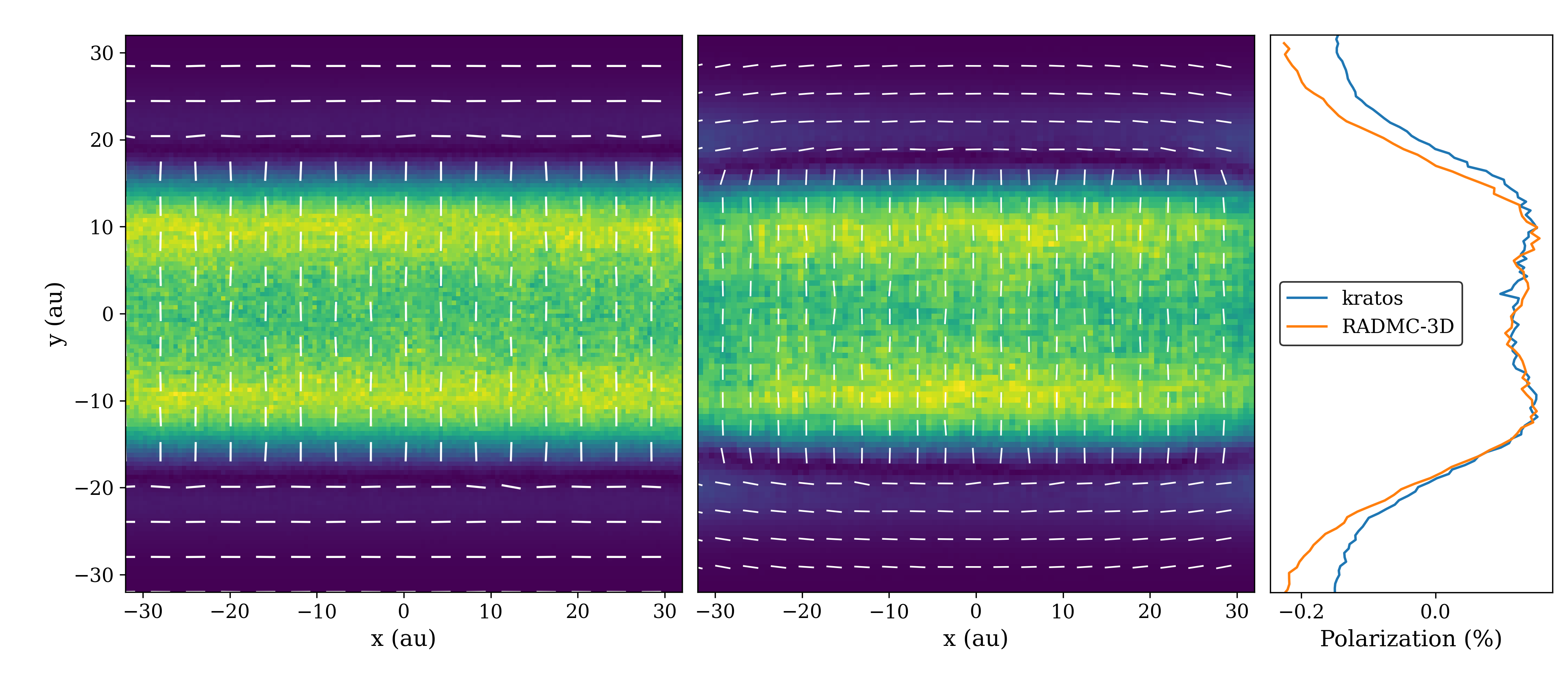}
  \caption{Comparison of self-scattering polarization in a
    dust cylinder. Left: \kratos{} simulation showing
    polarized intensity (heatmap) and orientation (line
    segments). Middle: Equivalent RADMC-3D
    results. Right: Polarization fraction profiles
    along $x=0$ demonstrate code agreement, with boundary
    differences attributable to periodic conditions in
    \kratos{}.}
  \label{fig:kataoka_img}
\end{figure*}

The second test addresses the critical balance between
scattered radiation and direct thermal emission during
imaging synthesis. This balance is non-trivial as it
requires proper treatment of both source terms during ray
tracing. For spherical dust grains, direct thermal emission
dilutes polarization signals, while for aligned irregular
grains, the interplay becomes more complex due to additional
polarization from emission.  We configured a cylindrical
dust distribution with isothermal temperature $T = 30$ K and
Gaussian density profile
$\rho(r) = \rho_0 \exp[-(r/r_0)^2]$, similar to the test in
\citep{2015ApJ...809...78K}. The dust composition follows
astronomical silicate \citep{2003ApJ...598.1017D}
with an
MRN size distribution \citep{1977ApJ...217..425M} and
maximum grain size $a_{\rm max} = 14\,\mu$m, ensuring
validity of the dipole approximation currently implemented
in \kratos{}, at the assumed wavelength of 870 $\mu $m.  
Figure~\ref{fig:kataoka_img} presents
polarized intensity maps with overlaid polarization
vectors. The right panel compares polarization fractions
(defined as $Q/I$) along the $x=0$ cut between \kratos{} and
RADMC-3D. The excellent agreement in polarization fraction
validates our implementation, while differences near
boundaries arise from our use of periodic boundary
conditions, which eliminate edge artifacts towards
x-boundaries present in the RADMC-3D simulation, as well as
reducing the parallel polarization towards y-boundaries.

\begin{figure}
  \centering
  \includegraphics[width=\linewidth]{\figdir/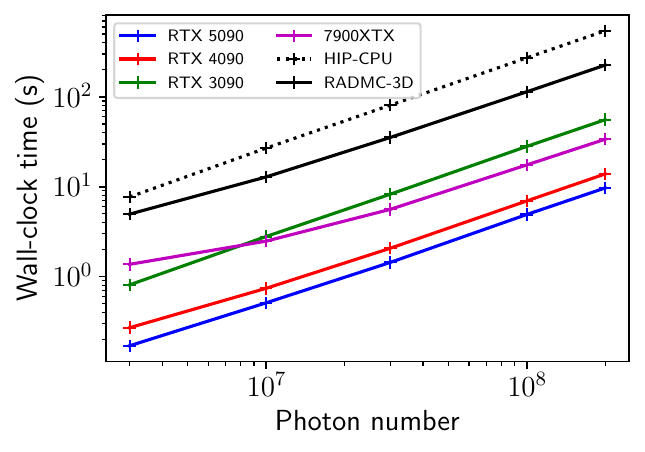}
  \caption{Speed test using the same physical layout
    described in \S\ref{sec:self-scattering}, showing the
    performance of \kratos{} with varying numbers of
    simulated photon packets compared to RADMC-3D. The
    HIP-CPU case conducts \kratos{} calculations using the
    AMD Ryzen 7950X CPU, which is the same device used by
    the RADMC-3D tests. Other tests using RTX devices and
    7900XTX are conducted on GPUs.}
  \label{fig:spd_tst}
\end{figure}

Based on the same test problem, we benchmarked computational
performance against RADMC-3D, with results shown in
Figure~\ref{fig:spd_tst}. The GPU-accelerated \kratos{}
demonstrates superior scaling, with computation time growing
approximately linearly with photon number. On both NVIDIA
(RTX series) and AMD (7900XTX) hardware, \kratos{} exceeds
CPU performance by over an order of magnitude.  Notably,
\kratos{} maintains this performance advantage while
including polarized absorption processes that RADMC-3D
simplifies. The RTX 3090 achieves throughput equivalent to
$\sim 80$ CPU cores of AMD Ryzen 7950X, while the RTX 5090
matches $\sim 320$ CPU cores, highlighting the
transformative potential of GPU acceleration for polarized
radiative transfer calculations. For the runs across the
different platforms on which we have performed tests, the computed
Stokes parameters are largely consistent. Users are advised that,
under certain versions of GPU drivers and RNG implementations, the
quality of the random numbers may lead to artifacts in the outputs
such as barely visible strips in the image plots. Updates of the
driver or switching to different devices could likely resolve such
issues. We also note that the impact of these effects is minimal on
the scientific outputs.

\subsection{Polarization Rotation in Twisted Magnetic Fields}
\label{sec:twisted-bfield-test}

\begin{figure}
  \centering
  \includegraphics[width=\linewidth]{\figdir/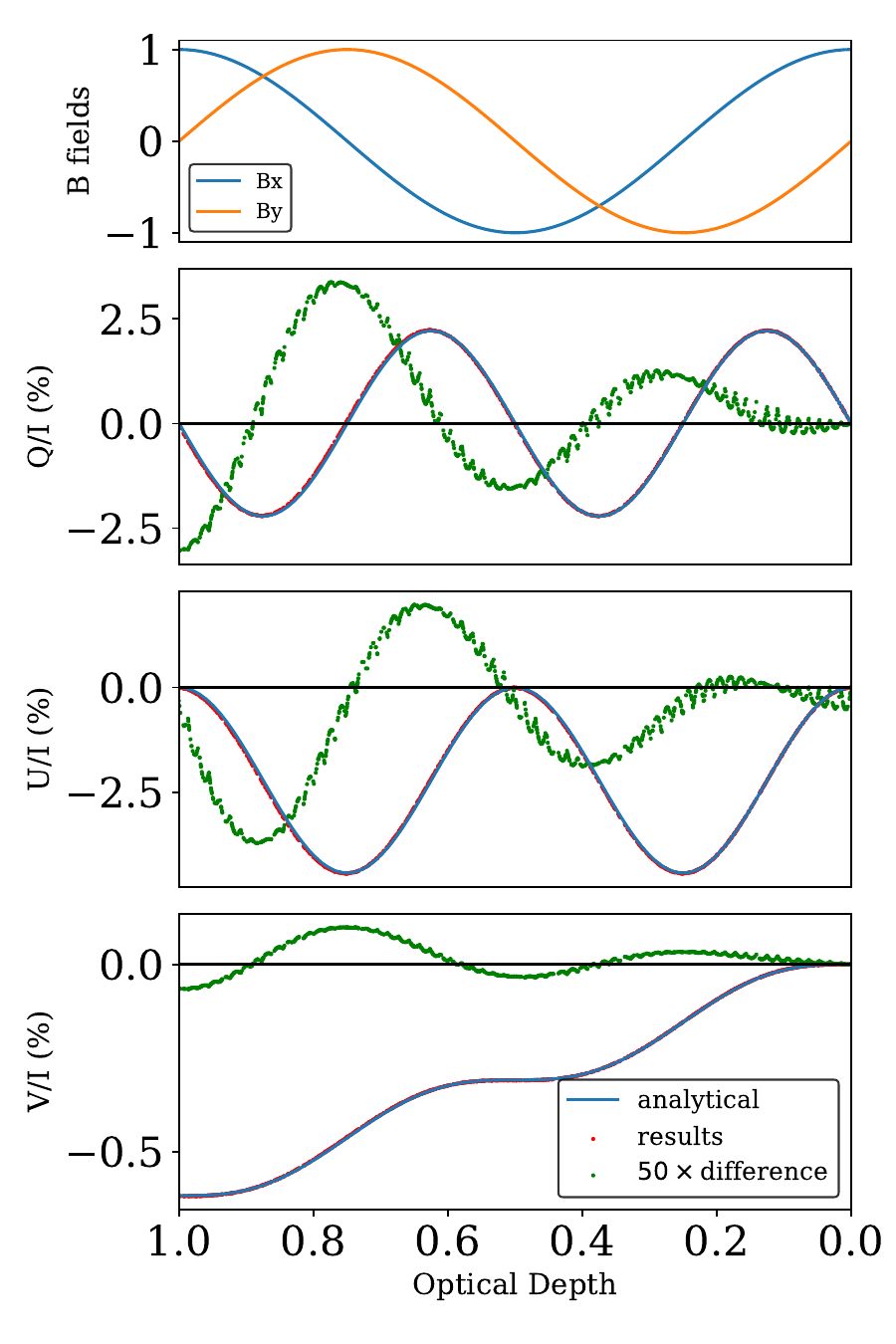}
  \caption{Emergent polarization from a slab with twisting
    magnetic fields. Top panel: Magnetic field
    configuration showing continuous rotation across the
    slab. Lower panels: Normalized Stokes
    parameters $Q/I$, $U/I$, and $V/I$ as functions of
    generation optical depth. Simulation results (red
    dots) show excellent agreement with analytical
    predictions (blue lines). Residual differences (green
    dots, magnified 50$\times$) demonstrate the accuracy
    of the polarization transformation implementation.}
  \label{fig:btwist}
\end{figure}

To validate the accurate treatment of polarization state
transformations during propagation through magnetically
aligned media, we designed a test scenario involving light
propagation through a slab with spatially varying magnetic
field orientations. This test specifically examines the
rotation behavior of Stokes parameters as photons traverse
regions with continuously changing grain alignment
directions. We constructed a simulation domain consisting of
a slab with total optical depth $\tau_{\rm total} = 1$
embedded with a twisting magnetic field configuration. The
magnetic field vector varies sinusoidally throughout the
slab with zero $z$-component and spatially varying $x$- and
$y$-components:
\begin{equation}
  \label{eq:bfield-config}
  B_x(\tau) = B_0 \cos(2\pi\tau), \quad B_y(\tau) = B_0
  \sin(2\pi\tau)\ ,
\end{equation}
where $\tau$ represents the optical depth coordinate through
the slab. This configuration creates a magnetic field that
rotates continuously by $360^\circ$ across the slab
thickness, as illustrated in the top panel of
Figure~\ref{fig:btwist}.

We initialized $10^3$ unpolarized photons distributed
randomly throughout the slab volume. These photons propagate
vertically through the medium until exiting the simulation
domain. During propagation, they experience dichroic
extinction due to dust grains aligned with the local
magnetic field direction, with each interaction inducing
polarization dependent on the grain orientation relative to
the photon's propagation direction and polarization state.

In the optically thin limit, where polarization from
secondary extinction can be neglected, the emergent
polarization can be derived analytically by integrating the
contributions from successive layers with varying magnetic
field orientations, each operating on their own
non-polarized photons.  For initially unpolarized light
passing through the twisting field configuration, the
analytical expressions for the normalized Stokes parameters
are,
\begin{equation}
\label{eq:analytical-qu}
\left\{
\begin{split}
\frac{Q}{I} &= \frac{p_0}{4\pi} \sin(4\pi\tau) \\
\frac{U}{I} &= -\frac{p_0}{4\pi}[1 - \cos(4\pi\tau)] \\
\frac{V}{I} &= \left(\frac{p_0}{4\pi}\right)^2
              [\sin(4\pi\tau) - 4\pi\tau]
\end{split}
\right. ,
\end{equation}
where $p_0$ represents the intrinsic polarization fraction
in the optically thin limit, determined by the dielectric
function and aspect ratio of the dust grains.

The extinction matrix, calculated using the dipole
approximation and optical theorem, contains no phase lag
between differently polarized dipole
excitations. Consequently, there are no direct cross-terms
between Stokes $U$ and $V$ in the physical extinction
process. To test the code's handling of potential
cross-terms, we artificially introduced a $U$-$V$ coupling
term identical to the $I$-$Q$ cross-term in the extinction
matrix. This modification enables the generation of
appreciable circular polarization from initially unpolarized
light through the twisting field configuration, even in the
dipole approximation regime, providing a stringent test of
the coordinate transformation implementation.

Figure~\ref{fig:btwist} presents the comparison between
simulation results and analytical predictions. The emergent
polarization states of photons, plotted as functions of
their generation optical depth $\tau$, show excellent
agreement with the analytical solutions. The simulation data
(red dots) closely follow the theoretical curves (blue
lines) for all Stokes parameters. The residual differences,
magnified by a factor of 50 and shown as green dots, reveal
two distinct patterns: small-scale fluctuations attributable
to numerical errors within individual grid cells, and
larger-scale variations arising from the optical thin
approximation used in deriving the analytical
solutions. Despite these minor discrepancies, the overall
agreement demonstrates the correctness of the
quaternion-based coordinate transformation implementation in
handling the complex polarization state rotations through
media with spatially varying alignment directions.

\subsection{Validation of Dichroic Extinction by Aligned Grains}
\label{sec:extinction-test}

The accurate treatment of dichroic thermal emission and
extinction by aligned, non-spherical dust grains represents
a fundamental capability for simulating polarized radiation
in magnetized environments. To validate this critical
component, we examine the polarization behavior arising from
pure dichroic extinction in a uniformly aligned dust slab
without scattering contributions.  The theoretical
polarization fraction for such a system follows the
analytical expression derived by
\citet{2000PASP..112.1215H}:
\begin{equation}
  \label{eq:ptau}
  P = \dfrac{\e^{-\tau}\sinh(p_0 \tau)}
  {1 - \e^{-\tau}\cosh(p_0 \tau)},
\end{equation}
where $\tau$ denotes the optical depth and $p_0$ represents
the intrinsic polarization determined by grain aspect ratio,
dielectric properties, and viewing angle relative to the
grain alignment axis.

\begin{figure}
  \centering
  \includegraphics[width=\linewidth]{\figdir/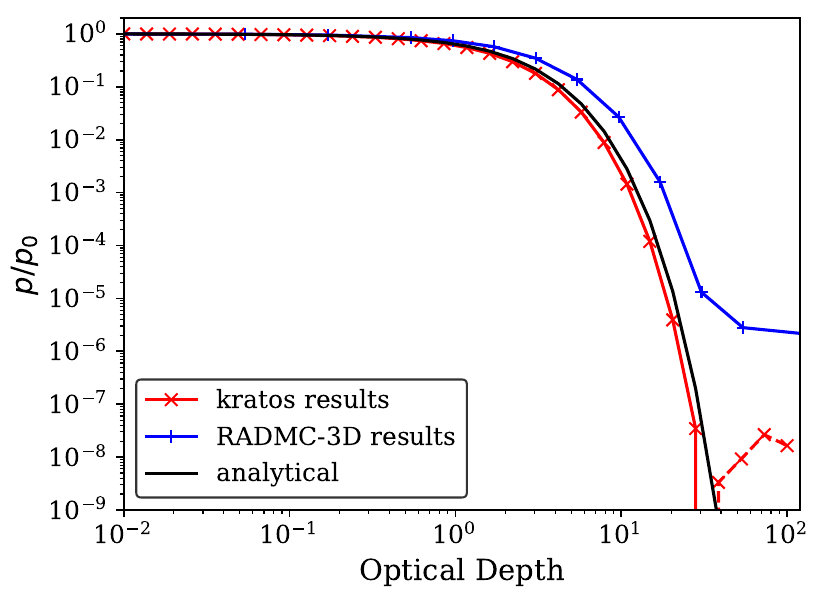}
  \caption{Polarization fraction as a function of optical
    depth in a uniformly aligned dust slab. Simulation
    results from \kratos{} (red) show excellent agreement
    with analytical theory (black) up to $\tau \approx 30$,
    outperforming RADMC-3D (blue) in the high-optical-depth
    regime.  The polarization is reversed at high optical
    depth in \kratos{} results, shown as dashed lines.  }
  \label{fig:ptau}
\end{figure}

Figure~\ref{fig:ptau} compares our simulation results (red
curves) with this analytical solution (black curve). The
\kratos{} implementation demonstrates excellent agreement
with theory up to $\tau \approx 30$, beyond which numerical
truncation errors become significant as polarization signals
approach the noise floor. We further compare our results
with RADMC-3D, which also implements dichroic processes. As
shown by the blue curve in Figure~\ref{fig:ptau}, RADMC-3D
begins to deviate from the theoretical solution at
$\tau \sim 3$, earlier than \kratos{}. This improved
performance stems from our implementation of analytical
solutions to the extinction matrix within each computational
cell, enhancing numerical stability in optically thick
regimes. Nevertheless, these differences remain negligible
for practical applications since the polarization fraction
from dichroic processes is usually below observational
limits at this high optical depth.

\subsection{Scattering by Grains in Plane-parallel Atmospheres}
\label{sec:plane-parallel-test}

To validate polarized scattering in more complex
environments, we simulate radiation transfer through a
plane-parallel atmosphere containing aligned dust grains,
following the methodology of \citet{Lin2022}. The uniform
slab incorporates dust grains with aspect ratio $s = 1.03$
and magnetic field alignment along the x-axis, creating
conditions where scattering and emission processes interact
significantly at moderate optical depths. We fix the
line-of-sight inclination at $45^\circ$ while varying the
azimuthal viewing angle to probe the full polarization
response.

\begin{figure*}
  \centering
  \includegraphics[width=1.05\textwidth]{\figdir/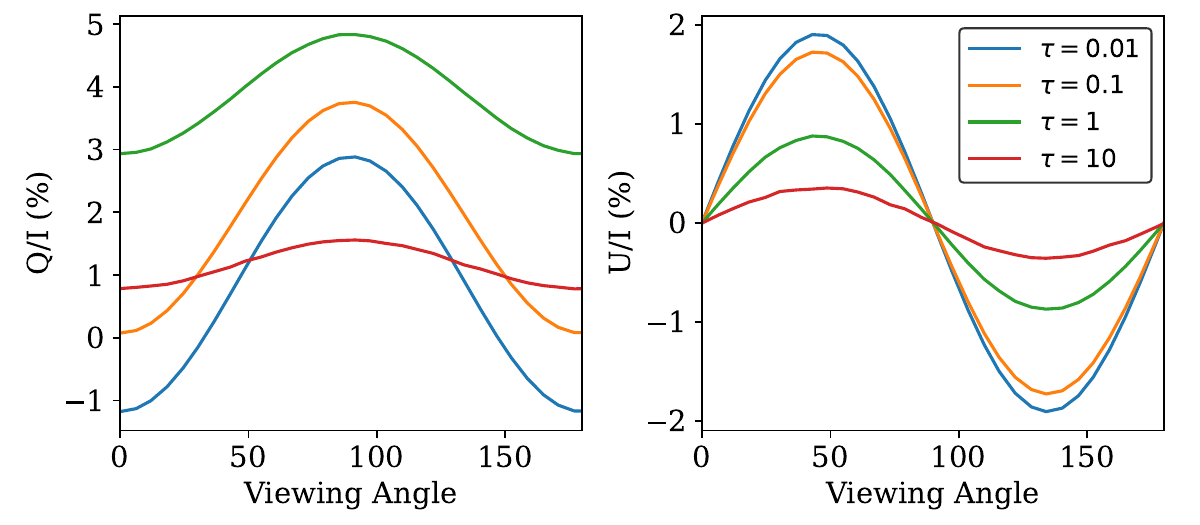}
  \caption{Polarization patterns in plane-parallel
    atmospheres with aligned dust grains. Stokes
    parameters $Q/I$ and $U/I$ are shown as functions of
    azimuthal viewing angle at fixed $45^\circ$
    inclination for different optical depths. Results
    demonstrate the interplay between direct emission
    modulation and scattering-induced polarization,
    showing excellent agreement with established
    benchmarks.}
  \label{fig:ppatm}
\end{figure*}

Figure~\ref{fig:ppatm} presents the resulting Stokes
parameters $Q/I$ and $U/I$ as functions of viewing angle for
different optical depths.  The polarization patterns reveal
two distinct components: a modulated contribution from
direct thermal emission that varies with viewing geometry
relative to grain orientation, and a scattering-induced
component that produces a non-zero offset in Stokes $Q$. The
emission modulation diminishes with increasing optical
depth, consistent with the damping behavior described by
Equation~\ref{eq:ptau}.  The Stokes parameter $U$ oscillates
around zero as expected for this symmetric configuration,
while Stokes $Q$ exhibits a characteristic offset from
self-scattering polarization. This offset displays
non-monotonic dependence on optical depth, initially
increasing until $\tau \sim 1$ before declining to an
asymptotic value of approximately $1\%$, in agreement with
the findings of \citet{Yang2017}. These results demonstrate
excellent consistency with established benchmarks,
validating our implementation of polarized scattering by
aligned grains.

\subsection{Scattering of Central Starlight by Aligned Grains}
\label{sec:scat-central-star}

To validate our implementation against established
benchmarks, we reproduce the test case introduced by
\citet{2002ApJ...574..205W} in their HOCHUNK radiative
transfer code. This test examines polarized scattering by
magnetically aligned, irregular dust grains in a
circumstellar environment, using a uniform spherical
distribution as a computationally tractable yet physically
relevant configuration. We configure a uniform sphere of
radius 20 au with a homogeneous magnetic field oriented
along the z-axis. Dust grains are modeled as perfect oblate
spheroids with short axes aligned parallel to the magnetic
field direction. For the grain properties, we adopt a
complex refractive index $m = 2.30 + 0.0228{\rm i}$ and
aspect ratio 2:1, similar to the implementation of
\citep{2002ApJ...574..205W}.

\begin{figure*}
 \centering
 \includegraphics[width=0.95\textwidth]{\figdir/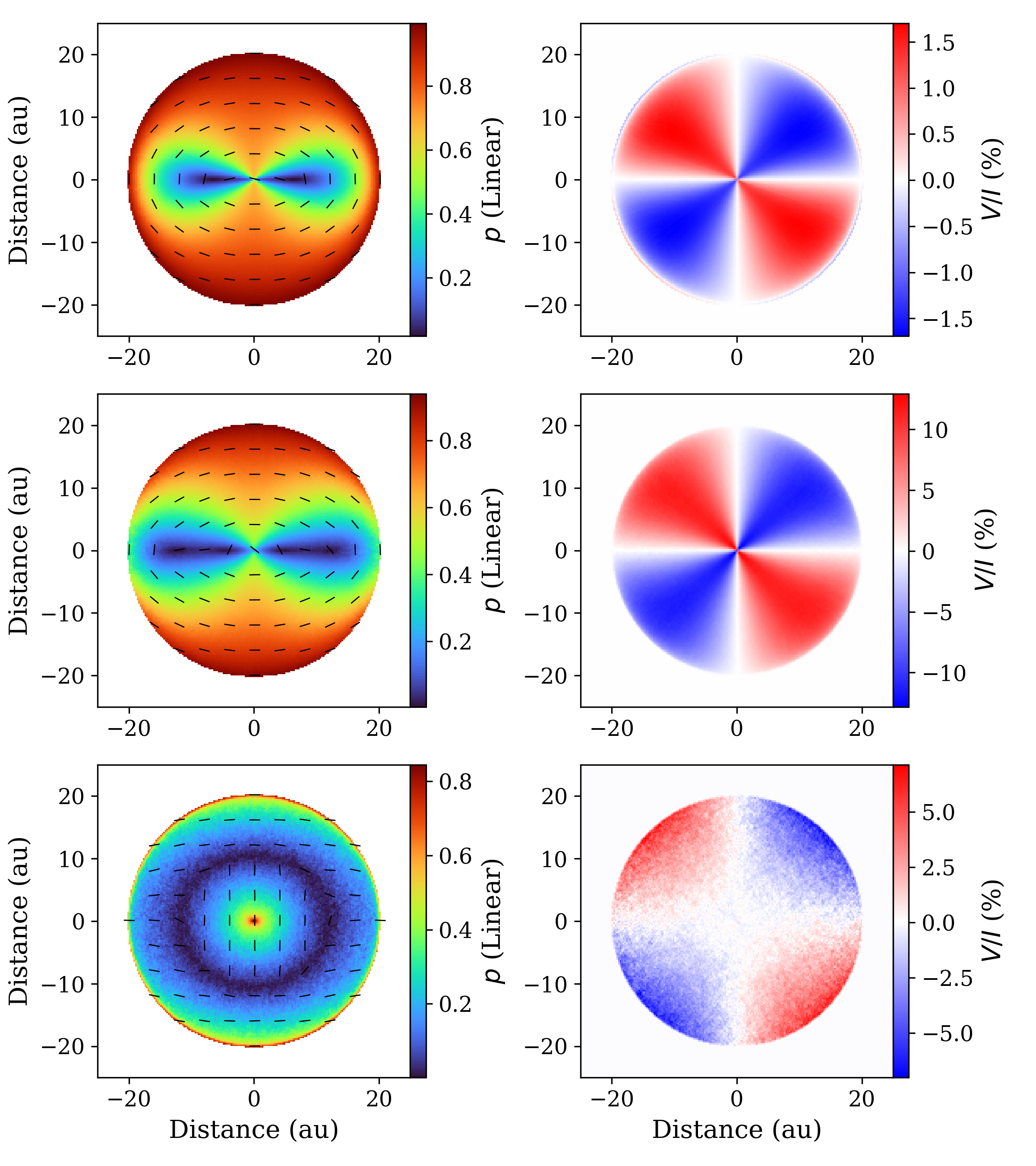}
 \caption{Polarization patterns from scattering by aligned
   dust grains in a uniform sphere. Top to bottom:
   Results for optical depths $\tau = 0.01$, 1, and 10.
     Left column: Linear polarization fraction with
   overlaid orientation vectors. Right column:
   Circular polarization fraction. The transition from
   scattering-dominated to extinction-dominated regimes is
   clearly visible with increasing optical depth.}
  \label{fig:ww2002}
\end{figure*}

Figure~\ref{fig:ww2002} presents the resulting polarization
patterns for three characteristic optical depths
($\tau = 0.01$, 1, and 10), viewed along the x-axis
perpendicular to the magnetic field direction. In the
optically thin regime ($\tau = 0.01$), the polarization
pattern exhibits, to the leading order, concentric
modulation reflecting the central illuminating source.
Grains aligned with their long axes horizontally produce
preferential scattering that reduces polarization along this
direction, creating a distinctive figure-8 morphology in the
polarization map. As optical depth increases to $\tau = 1$,
this pattern expands while maintaining its fundamental
character, demonstrating the growing influence of
scattering. The optically thick case ($\tau = 10$) reveals a
qualitatively different pattern dominated by dichroic
extinction. Foreground grains preferentially absorb
radiation polarized along their long axes, resulting in
polarization aligned with the magnetic field direction in
the central regions. This transition from
scattering-dominated to extinction-dominated polarization
represents a critical validation of our code's ability to
handle the complex interplay between different polarization
mechanisms across optical depth regimes. Circular
polarization patterns remain qualitatively consistent across
optical depths, with lower polarization fractions in central
regions compared to outer areas. However, our implementation
produces systematically lower circular polarization
fractions than \citet{2002ApJ...574..205W}, which we
attribute to differences in the treatment of grain
optics. Our dipole approximation assumes minimal phase lag
between dipole excitations along different grain axes,
relying primarily on the imaginary component of the
dielectric function to generate Stokes $I$-$V$ coupling. The
predominantly refractory nature of our dust model
($\Im{m} \ll \Re{m}$) further suppresses circular
polarization generation compared to more absorptive grain
compositions.

\subsection{Convergence of the Stokes parameters}
\label{sec:convergence}

As practical guidance for users of the code, we characterize how
the accuracy of the computed Stokes parameters depends on the number of
photon packets. We adopt a uniform, isothermal slab of dust with optical
depth $\tau$, whose grains have a single-scattering albedo $\omega$, and
view it at an inclination of $45^\circ$. Because the slab is uniform and the
polarization is produced by self-scattering, the emerging signal is carried
by Stokes $Q$ alone ($U \simeq 0$); the scattered-light image is uniform by
construction, so we take the per-pixel noise to be the dispersion of the
Stokes values across the image pixels.

\begin{figure*}
  \centering
  \includegraphics[width=\linewidth]{\figdir/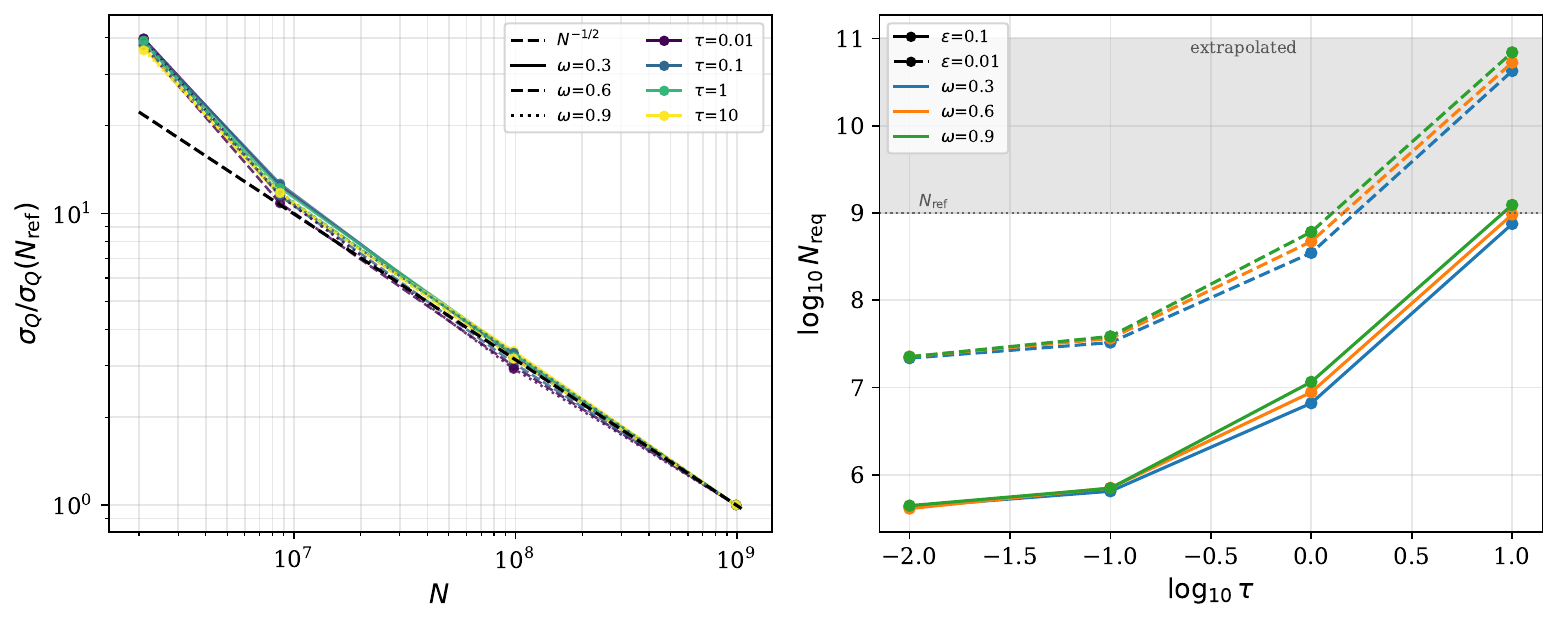}
  \caption{Convergence of the Stokes parameters in the
    uniform-slab test. Left: per-pixel dispersion of Stokes $Q$,
    normalised to its value at the reference run $N_{\rm
    ref}=9.9\times10^{8}$, versus packet number $N$ for the optical
    depths $\tau$ (colours) and albedos $\omega$ (line styles); the 12
    curves shown all collapse onto the Poisson $N^{-1/2}$ scaling
    (dashed). Right: number of packets $N_{\rm req}$ required to reach a
    fractional accuracy $\varepsilon$ in Stokes $Q$ as a function of
    optical depth, for three albedos. The horizontal dotted line
    and grey shading mark $N_{\rm ref}=9.9\times10^{8}$, the largest
    packet number actually simulated; points above it are extrapolations
    of the measured $N^{-1/2}$ scaling, not additional runs. The
    required photon number is governed primarily by the optical depth
    and only weakly by the albedo.}
  \label{fig:convergence}
\end{figure*}

The result is shown in Fig.~\ref{fig:convergence}. Relative to the
highest-photon-number reference run, the per-pixel dispersion of Stokes $Q$
follows the expected Monte Carlo scaling $\sigma_Q \propto N^{-1/2}$ for all
optical depths and albedos considered, confirming that the polarized
two-step estimator is limited only by Poisson noise. The absolute level,
however, is problem-dependent: the number of packets $N_{\rm req}$ required
to reach a given fractional accuracy $\varepsilon$ grows rapidly with
optical depth and only weakly with albedo, so no single photon number
applies to all configurations. The values of $N_{\rm req}$ in the
right-hand panel are extrapolations of the measured $N^{-1/2}$ scaling from
the largest simulated run, $N_{\rm ref}=9.9\times10^{8}$; the grey region
above $N_{\rm ref}$ marks this extrapolated range. For reference, the
photon numbers indicated here are consistent with the
$\sim 5\times10^{9}$ packets used by \citet{Yang2020} to obtain a low-noise
synthetic polarization image of a T Tauri disk with \textsc{radmc-3d}.

We emphasize that this slab test is illustrative. Real synthetic
observations are usually convolved with the beam of the instrument, which
averages over neighbouring pixels and reduces the noise relative to the raw
image, so the photon number actually required depends on the target
accuracy, the beam size, and the physical set-up. We therefore encourage
users to perform a convergence test tailored to their own problem.

\section{Discussion}
\label{sec:discussion}

\subsection{Astrophysical implications}

The self-consistent treatment of scattering by aligned grains
presented in this work enables observational diagnostics that
are inaccessible to spherical-grain MCRT codes.  At
ALMA wavelengths, the polarization characteristics of
protoplanetary disks encode information about both the dust
population and the underlying alignment mechanism.  In the
spherical-grain approximation, scattering produces an
axially symmetric azimuthal polarization pattern whose
magnitude is governed solely by the scattering angle,
optical depth, and grain albedo
\citep{Yang2016a}.  Adding aligned grains breaks this axial
symmetry: the B-field orientation imprints an angular
modulation on the Stokes parameters (Figs.~\ref{fig:ppatm}
and \ref{fig:ww2002}) that spherical-grain codes,including
RADMC-3D, MCFOST, SKIRT, and Hyperion,cannot reproduce.
POLARIS includes dichroic emission from aligned grains but
does not couple grain alignment to the scattering phase
function, so it is likewise unable to model the
self-consistent transition from scattering-dominated (Band
7) to dichroic-dominated (Band 3) polarization that has been
observed in disks such as HL Tau
\citep{2015ApJ...809...78K}.  Owing to this limitation,
existing studies of the interplay between dichroic emission
and scattering have been forced to rely on approximations:
either computing the two processes with separate codes and
adding the resulting Stokes parameters linearly \citep{2021ApJ...908..153M}, or adopting a plane-parallel slab model
\citep{Lin2022} and extending the solution to global disk
geometries \citep{2023Natur.623..705S, Lin2024a}.  While
pragmatic, both approaches neglect the coupled evolution of
the polarization state when grains simultaneously scatter
and emit within the same medium.  \kratos{} is, to our
knowledge, the only modern publicly maintained code capable
of performing self-consistent calculations of this coupled
process.

\subsection{Limitations}

Several approximations in the current implementation warrant
discussion.  First, the dipole approximation for grain
optical properties is valid only when the grain size
parameter $x = 2\pi a / \lambda \lesssim 1$.  At ALMA
Band 3 ($\lambda \approx 3\,$mm), this condition holds for
grains up to $a \sim 500\,\mu$m, covering the typical range
for disk dust populations; at Band 7 ($\lambda \approx
870\,\mu$m) it is marginally satisfied for $a \lesssim
140\,\mu$m.  For larger grains, Mie theory or T-matrix
calculations become necessary, and we plan to address this
through neural-network emulators (see below).  We note,
however, that the radiative transfer infrastructure
described in \S\ref{sec:methods} is agnostic to the grain
model: the Mueller matrix $Z(\Omega; \Omega')$ can be
supplied by any method, so the code itself is ready for more
sophisticated grain optics as they become available.

Second, we have assumed perfect alignment of grain short
axes with the local magnetic field.  In reality, grain
alignment is imperfect and varies with grain size, shape,
and the local radiation field; partial alignment reduces
polarization amplitudes without fundamentally changing the
morphology.  Incomplete alignment can be incorporated
through a reduction factor applied to the polarization cross
sections \citep{2015ARA&A..53..501A}, and we will add this
capability in a future release.

Third, the present verifications are restricted to
analytical solutions and controlled benchmark problems.  We
have not performed direct fits to observed polarization maps
of specific sources; such application work requires tuning
the dust model, disk structure, and alignment configuration
and is beyond the scope of this methodology paper.  We defer
systematic application to observed disks to future work.

\subsection{Future directions}

Two astrophysical applications are particularly motivated by
the results presented here.  The first is self-consistent
simulations of the ALMA polarization transition across
Bands 7, 6, and 3 for sources such as HL Tau, where the
relative contributions of self-scattering and dichroic
emission can be separated for the first time, enabling
direct constraints on grain sizes and alignment
mechanisms.  The second is NIR polarimetry of disk
surfaces, where distinguishing the $U_\phi$ signal produced
by B-field-aligned scattering from that produced by
multiple scattering of spherical grains requires the
self-consistent treatment provided by \kratos{}
\citep{Yang2022}.

On the code-development side, the most promising extension
is the integration of neural-network emulators for complex
dust grain optical properties.  Realistic dust models must
account for distributions in grain sizes, shapes (including
fractal dimensions and filling factors), compositions, and
alignment efficiencies, producing high-dimensional functions
mapping incident photon configurations to scattered
properties.  High-dimensional interpolation tables are
memory-prohibitive for GPU Monte Carlo; neural-network
surrogates trained on T-matrix or DDA calculations offer a
compact alternative that would dramatically expand the
complexity of dust models accessible to polarized MCRT
simulations.

\section{Summary}
\label{sec:summary}

We have presented \kratos{}, a GPU-accelerated Monte Carlo
radiative transfer code that provides the first
self-consistent treatment of scattering by aligned,
non-spherical dust grains in a modern, publicly maintained
framework.  The code employs quaternion-based coordinate
transformations,validated by $10^7$ randomized round-trip
tests to machine precision,and analytical cell-by-cell
integration of the polarized transfer equation for numerical
stability at high optical depth.  Extensive verification
against analytical solutions and established codes
demonstrates the accuracy of the implementation across
diverse polarization phenomena, including
self-scattering, dichroic extinction, and the
scattering--extinction transition in aligned-grain
media.  The code achieves a speedup of $\sim 10^{2}$
relative to CPU-based methods, making parameter-space
exploration and high-resolution imaging synthesis practical
on consumer-grade GPUs.  \kratos{} is publicly available
and constitutes a necessary tool for modeling polarized
radiation in systems where large grains both scatter
strongly and align with ambient fields.

\section*{Acknowledgements}

We thank the anonymous referee for constructive comments that
helped improve the manuscript.

This work is supported by the National Natural Science
Foundation of China (NSFC) [12473067].

During the preparation of this manuscript, the authors made use of
the large language model DeepSeek V4 Pro for tasks
including structural planning, copyediting, and text refinement. All
suggestions and AI-generated content were critically reviewed,
verified for scientific accuracy, and edited by the authors, who take
full responsibility for the final manuscript in accordance with the
COPE position statement on Authorship and AI.

\section*{Data Availability}

The \kratos{} source code is publicly available at
\url{https://github.com/wll745881210/kratos_polrad.git}.
Simulation data and analysis scripts are available from the
corresponding author upon reasonable request.

\appendix

\section{Solution to radiative transfer}
\label{sec:apdx-ana-pol-rt}

Once the vector source function per extinction optical depth
$(S_1, S_2, S_3, S_4)$ (including the emission source
function, and the scattering source function if applied in
the imaging step; see also \S\ref{sec:method-rt}) has been
determined, we recast the radiative transfer equation
(eq.~\ref{eq:polarized-rt}) in the following expanded form,
\begin{equation}
  \dfrac{\d}{\d\tau}
  \begin{bmatrix} I \\ Q \\ U \\ V \end{bmatrix}
  = -
  \begin{bmatrix} \alpha_1 & \alpha_2 & 0 & 0 \\
    \alpha_2 & \alpha_1 & 0 & 0 \\
    0 & 0 & \alpha_1 & \alpha_3 \\
    0 & 0 & -\alpha_3 & \alpha_1
  \end{bmatrix}
  \begin{bmatrix} I \\ Q \\ U \\ V \end{bmatrix}
  +
  \begin{bmatrix} S_1 \\ S_2 \\ S_3 \\ S_4 \end{bmatrix},
\end{equation}
where $\d \tau$ denotes the differential extinction optical
depth (combining absorption and scattering) along the
propagation path of a photon packet.  The matrix elements
$\alpha_i$ ($i=1,2,3$) are determined based on dust grain
properties (in case of dipole grain approximation, for
example, see also the appendix of \citealt{Yang2022}).  Let
$(I_0, Q_0, U_0, V_0)$ be the Stokes parameters entering
this cell, and $d\tau$ be the extinction optical depth of
the light trans-passing this cell. With extinction optical
depth $\delta \tau$ (which could be optically thick)
measured in one cell from one cell interface to another, the
Stokes parameters of the light after propagating through the
cell can be solved analytically,
\begin{equation}
  \begin{split}
    I = I_{s} &+ (I_{0}-I_{s}) \e^{-\alpha_{1}\delta
                \tau} \cosh(-\alpha_{2}\delta
                \tau)\\
    &+ (Q_{0}-Q_{s}) \e^{-\alpha_{1}\delta \tau}
    \sinh(-\alpha_{2}\delta \tau)\ ,\\
    Q = Q_{s} &+ (I_{0}-I_{s}) \e^{-\alpha_{1}\delta
                \tau} \sinh(-\alpha_{2}\delta
                \tau) \\
    &+ (Q_{0}-Q_{s}) \e^{-\alpha_{1}\delta \tau}
    \cosh(-\alpha_{2}\delta \tau)\ ,\\
    U = U_{s} &+ (U_{0}-U_{s}) \e^{-\alpha_{1}\delta \tau}
    \cos(-\alpha_{3}\delta \tau)\\
    &+ (V_{0}-V_{s}) \e^{-\alpha_{1}\delta \tau}
    \sin(-\alpha_{3}\delta \tau)\ ,\\
    V = V_{s} &- (U_{0}-U_{s}) \e^{-\alpha_{1}\delta \tau}
    \sin(-\alpha_{3}\delta \tau)\\
    &+ (V_{0}-V_{s}) \e^{-\alpha_{1}\delta \tau}
    \cos(-\alpha_{3}\delta \tau)\ ,
  \end{split}
\end{equation}
where the auxiliary quantities are defined as,
\begin{equation}
\begin{split}
I_s &= \frac{\alpha_1S_1-\alpha_2S_2}{\alpha_1^2-\alpha_2^2} \\
Q_s &= \frac{\alpha_1S_2-\alpha_2S_1}{\alpha_1^2-\alpha_2^2} \\
U_s &= \frac{\alpha_1S_3-\alpha_3S_4}{\alpha_1^2+\alpha_3^2} \\
V_s &= \frac{\alpha_1S_4+\alpha_3S_3}{\alpha_1^2+\alpha_3^2}
\end{split}
\end{equation}

\bsp
\label{lastpage}
\end{document}